\documentclass{ws-ijmpd}
\usepackage[super,compress]{cite}
\begin{document}

\markboth{Jos\'e P. S. Lemos, Francisco J. Lopes, Masato Minamitsuji}
{Rotating thin shells in (2+1)-dimensional spacetimes}

%
\catchline{}{}{}{}{}
%

\title{ROTATING THIN SHELLS IN (2+1)-DIMENSIONAL 
ASYMPTOTICALLY ADS SPACETIMES: MECHANICAL PROPERTIES, MACHIAN EFFECTS, 
AND ENERGY CONDITIONS}
\author{JOS\'E P. S. LEMOS}

\address{Centro Multidisciplinar de Astrof\'isica - CENTRA,
Departamento de F\'isica, Instituto Superior T\'ecnico - IST,
Universidade de Lisboa - UL, Av. Rovisco Pais 1\\ Lisboa, 1049-001,
Portugal\\ joselemos@tecnico.ulisboa.pt}

\author{FRANCISCO J. LOPES}

\address{Centro Multidisciplinar de Astrof\'isica - CENTRA,
Departamento de F\'isica, Instituto Superior T\'ecnico - IST,
Universidade de Lisboa - UL, Av. Rovisco Pais 1\\ Lisboa, 1049-001,
Portugal\\ franciscojoaolopes@tecnico.ulisboa.pt}

\author{MASATO MINAMITSUJI}

\address{Centro Multidisciplinar de Astrof\'isica - CENTRA,
Departamento de F\'isica, Instituto Superior T\'ecnico - IST,
Universidade de Lisboa - UL, Av. Rovisco Pais 1\\ Lisboa, 1049-001,
Portugal\\ masato.minamitsuji@tecnico.ulisboa.pt}

\maketitle

\begin{abstract}
A rotating thin shell in a (2+1)-dimensional asymptotically AdS
spacetime is studied.  The  spacetime exterior to the shell is the
rotating BTZ spacetime and the interior is the empty spacetime with a
cosmological constant. Through the Einstein equation in
(2+1)-dimensions and the corresponding junction conditions we
calculate the dynamical relevant quantities, namely, the rest
energy-density, the pressure, and the angular momentum flux density. We
also analyze the matter in a frame where its energy-momentum tensor
has a perfect fluid form.  In addition, we show that Machian effects,
such as the dragging of inertial frames, also occur in rotating
(2+1)-dimensional spacetimes.  The weak and the dominant energy
condition for these shells are discussed.
\end{abstract}




\section{Introduction}

In (2+1)-dimensional general relativity with a cosmological constant,
spacetimes with no matter have no propagating degrees of freedom, so
that vacuum spacetimes have local constant curvature.  Nonetheless, in
the presence of a negative cosmological constant, there is a
nontrivial black hole solution, the Ba\~nados, Teitelboim, Zanelli
(BTZ) black hole \cite{btz1,btz2}. In its most general form the BTZ
black hole is a stationary solution representing a rotating black
hole, with mass and angular momentum.  For zero rotation the BTZ black
hole is static.  The BTZ black hole, having a negative cosmological
constant, is an asympotically anti-de Sitter (AdS) solution.

Besides being a pure vacuum solution of a spacetime with a negative
cosmological constant, the BTZ solution is also a solution exterior to
a (2+1)-dimensional matter configuration.  A simple matter
configuration is a thin shell.  Collapsing nonrotating thin shells
with an exterior static BTZ solution have been studied in
\cite{pelegsteif,mannohpark0}$\,$.  Static thin shells with an
exterior static BTZ solution were worked out in
\cite{LemosQuintaBTZ}$\,$.  Gravitational collapse and configurations
of rotating thin shells with an exterior stationary BTZ solution was
analyzed in \cite{mannohpark}$\,$.

Here, we want to study rotating thin shells for which the exterior
spacetime is the stationary BTZ solution and the interior spacetime is
the empty spacetime with a cosmological constant.  We translate to 2+1
dimensions the framework of Poisson to treat rotating shells in
(3+1)-dimensional general relativity \cite{Poisson}$\,$, and thus,
through the Einstein equation in (2+1)-dimensions and the
corresponding junction conditions, we calculate the dynamical relevant
quantities.

This paper is organized as follows. In Sec.~\ref{thinshellfeat} we
perform an analysis of the (2+1)-dimensional spacetime with a negative
cosmological constant generated by a rotating thin matter shell. The
exterior is the BTZ solution, the interior is the empty spacetime, and
the energy-density, the pressure, and the angular momentum flux
density of the matter in the shell are calculated. In
Sec.~\ref{perfectfluid} we analyze the thin shell's mechanical
properties in a frame where one can detect the energy-momentum tensor
as a perfect fluid. 
In Sec.~\ref{sevrotations}
we study rotating properties of the spacetime and
find that (2+1)-dimensional spacetimes can display Machian effects,
such as the dragging of inertial frames.  In
Sec.~\ref{energyconditions} we discuss the weak and dominant energy
conditions for the thin shell matter. We conclude in
Sec.~\ref{conclusions}. In an Appendix we rederive the
energy-momentum tensor in a perfect fluid form using a different
approach.

\section{The thin shell spacetime: BTZ outside, AdS inside}
\label{thinshellfeat}

\subsection{Einstein equation}
\label{ee}

In 2+1 dimensions, the Einstein equation is
\begin{equation}
G_{\alpha \beta}+\Lambda g_{\alpha \beta}=8\pi G\, T_{\alpha \beta}
\,, 
\label{ein} 
\end{equation} 
where $G_{\alpha \beta}$ is the Einstein tensor, 
$\Lambda$ is the
cosmological constant,
$g_{\alpha \beta}$ is the spacetime metric,
$G$ is the
gravitacional constant in 2+1 dimensions,
and $T_{\alpha \beta}$ is the energy-momentum
tensor of the matter fields.  We will choose
units with the velocity of light equal to one, so that $G$
has units of inverse of mass. Greek indices run from $0,1,2$,
with $0$ being the time index. Since the background
spacetime is AdS which has a negative cosmological constant, 
we define the AdS length $l$ through the equation 
\begin{equation}
-\Lambda = 1/l^2
\,.
\label{adslength} 
\end{equation} 

We consider a timelike shell with radius $R$ in a
(2+1)-dimensional spacetime. In two spatial
dimensions a shell is a ring. 
This ring divides spacetime into two
regions, an outer region and an inner region.

\subsection{The outside region}
\label{4sec:sectionbo}

The outside region is described by 
an exterior metric given by the BTZ line element \cite{btz1,btz2} 
\begin{align}\label{4::exterior}
ds_o^2 = & g_{o\,\alpha \beta} dx_o^\alpha dx_o^\beta =
 -\left(\dfrac{r^2}{l^2} - 8Gm+\frac{16G^2J^2}{r^2}\right)\,dt_o^2 
\nonumber\\
&+ \frac{dr^2}{\left(\dfrac{r^2}{l^2} - 8Gm + \dfrac{16G_{3}^2 J^2
}{r^2}\right)} 
+ r^2 \left(d\phi-\frac{4GJ}{r^2}dt_o\right)^2
\,,\quad r> R\,,&
\end{align}
written in a circularly symmetric 
outer coordinate system $x_o^\alpha = (t_o, r, \phi)$,
and where $m$ is the
Arnowitt-Deser-Misner (ADM) mass
and $J$ is the spacetime angular momentum.

\subsection{The inside region}
\label{4sec:sectionbi}

The interior region is the vacuum state
empty spacetime, a variant of pure AdS spacetime, with
metric given by \cite{btz1,btz2} 
\begin{align} \label{4::interior}
ds_i^2 = g_{i\,\alpha \beta} dx^\alpha dx^\beta = &
-\dfrac{\rho^2}{l^2} \,dt_i^2 + \dfrac{l^2}{\rho^2} d\rho^2 + \rho^2
d\psi^2 \,,\quad \rho<R\,,
\end{align}
written in a circularly symmetric  
inner coordinate system 
$x_i^\alpha = (t_i, \rho, \psi)$.

\subsection{The shell-ring region: the matching 
and junction conditions}
\label{sec:4sectiond}

Here we analyze the matching region, i.e., the ring.
From the outside, the metric 
at the shell $\Sigma$, $r=R$, is from Eq.~(\ref{4::exterior})
\begin{equation}\label{4::exteriorR}
ds^2_\Sigma= 
-\left(\dfrac{R^2}{l^2} - 8Gm+\frac{16G^2J^2}{R^2}\right)\,dt_o^2 
+ R^2 \left(d\phi-\frac{4G J}{R^2}dt_o\right)^2
,\quad r= R\,.
\end{equation}
Now, we want to remove
the off-diagonal term in the induced metric (\ref{4::exteriorR}), 
as viewed from the
outer region. For that
we go to a corotating frame
by defining a new polar coordinate 
\begin{equation}\label{newpsi}
\psi
= \phi - \Omega t_o\,.
\end{equation}
This makes the induced metric diagonal if
the angular velocity $\Omega$ is choosen to be
\begin{equation}
\Omega = \dfrac{4 G J}{R^2} \label{Omega} \, .
\end{equation}
Then, in coordinates $(t_o,\psi)$,
the induced metric at $\Sigma$ is written as
\begin{equation} \label{LEOia2}
ds_{\Sigma}^2 =  -\left(\dfrac{R^2}{l^2} - 8Gm + \dfrac{16 G^2J^2
}{R^2}\right)\,dt_o^2 + R^2 d\psi^2 \,,\quad r=R\,.
\end{equation}
From the inside, the induced metric
is obtained by setting $\rho=R$ in 
Eq.~\eqref{4::interior},
i.e., 
\begin{equation} \label{4::interiorR}
ds^2_\Sigma =
-\dfrac{R^2}{l^2} \,dt_i^2 +  R^2
d\psi^2  \,,\quad \rho=R\,.
\end{equation}
On the other hand, at the shell, the metric 
can be written in terms of its own proper 
coordinates $x_\Sigma^a=(\tau,\psi)$, 
with $\tau$ being the proper time of the shell.
So, the line element 
$ds^2_\Sigma =h_{ab}\,dx_\Sigma^adx_\Sigma^b$, with 
$h_{ab}$ being the intrinsic metric at the shell,
can be written as
\begin{equation} \label{4::R}
ds^2_\Sigma =
-d\tau^2 +  R^2
d\psi^2 \,.
\end{equation}

We have still to apply the first 
and second junction conditions, see, e.g.,
\cite{Poisson}$\,$. 
The first junction 
condition
states that the induced
metric on the shell 
must be the same on both sides of the shell
and at the shell. From Eqs.~(\ref{LEOia2})-(\ref{4::R})
this gives
\begin{equation}
d\tau^2
=
\left(\dfrac{R^2}{l^2} - 8Gm+ 
\dfrac{16 G^2\,J^2}{R^2}\right) dt_o^2
= \dfrac{R^2}{l^2} \,dt_i^2 \,.
\end{equation}
The second junction condition states that
\begin{equation}
S_{a b}= - \dfrac{1}{8 \pi G} \left( [K_{a b} ] - [K]
h_{ab} \right) \label{b::35} \,,
\end{equation}
where $S_{ab}$
is the energy-momentum tensor of the matter
in the shell,
$K_{ab}=n_{\alpha ; \beta} e_a^\alpha e_b^\beta$
is the second fundamental form of the shell, with
$n_\alpha$ being the normal vector to the shell, 
$e_a^\alpha$ are the tangent
vectors to the shell, 
the semicolon 
denotes covariant derivative, and a quantity
in square brackets denotes
the jump from the outside to the inside.
The tangent vectors $e_a^\alpha$ 
are such that 
$h_{ab}$ given through Eq.~(\ref{4::R})
can be written as the metric induced either from
the outside or the inside metric $g_{\alpha\beta}$ at $R$,
i.e.,  
$h_{ab}=g_{\alpha\beta}
e^\alpha_ae^\beta_b$, 
where $g_{\alpha\beta}$ stands to either
$g_{o\,\alpha \beta}$ or 
$g_{i\,\alpha \beta}$.
The three independent
components of the surface energy-momentum tensor 
can then be calculated using Eq.~(\ref{b::35})
to give
\begin{eqnarray}
S_\tau^\tau&=& -\dfrac{1}{8 \pi G l} \left(1 - \dfrac{l}{R} \sqrt{
\dfrac{R^2}{l^2} -
8Gm +  \dfrac{16 G^2J^2 }{R^2}}\right) \,,
\label{Stt}
\\
S_\psi^\psi &=& \dfrac{1}{8 \pi G l} \left( 
\dfrac{
R^4-16 G^2 l^2 J^2}
{l R^3  \sqrt{\dfrac{R^2}{l^2}  - 8Gm + \dfrac{16 G^2J^2
}{R^2}}} -1 \right )\,,
\label{psipsi}
\\
S_\psi^\tau &=& 
\frac{J}{2\pi R}
 \,.
\label{Stpsi}
\end{eqnarray}

\subsection{Properties of the matter of the ring shell}

These calculations were performed in the shell's rest frame
coordinate $(\tau,\psi)$. In this frame we 
define the proper frame quantities,
namely, 
the proper rest energy density $\lambda$,
the proper pressure $p$, and the 
proper angular momentum flux density $j$ 
as $\lambda=-S^\tau_\tau$, 
$p=S_\psi^\psi$ and 
$j=S_\psi^\tau$. So,
\begin{eqnarray}
\lambda &=& \dfrac{1}{8 \pi G l} \left(1 -
\dfrac{l}{R} \sqrt{ \dfrac{R^2}{l^2} - 8Gm + \dfrac{16 G^2J^2
}{R^2}}\right)
\,, \label{lambdafluid}
\\
p& =&\dfrac{1}{8 \pi G l} \left( -1 + 
\dfrac{
R^4-16G^2 l^2 J^2}
{l R^3  \sqrt{ \dfrac{R^2}{l^2} - 8Gm + \dfrac{16 G^2J^2
}{R^2}}} \right )\,, \label{pfluid}
\\
j &=& 
\dfrac{
J}{2 \pi 
R} \,.\label{flu}
\end{eqnarray}
If one considers slowly rotation the terms in
$J^2$ in Eqs.~(\ref{lambdafluid}) and (\ref{pfluid})
should be discarded, keeping the linear term in
Eq.~(\ref{flu}).

An interesting quantity is the total rest mass $M$ of the 
shell, given by 
\begin{equation}
M = 2 \pi R\lambda\,.
\label{Mrest}
\end{equation}
Then using Eqs.~(\ref{lambdafluid}) and (\ref{Mrest})
one gets an expression for the ADM mass-energy $m$ in terms of 
the other quantities,
\begin{equation}
m = \dfrac{R}{l}M-2G M^2+2G\dfrac{J^2}{R^2}
\,.
\label{menergy}
\end{equation}
This expression connects shell quantities, namely, 
$M$ and $R$ with spacetime
quantities, namely, $m$, $J$ and $l$.

\subsection{Properties of the spacetime}

Inside the shell the spacetime is empty spacetime.
The matching region is the ring-shell region. Outside 
the shell the spacetime is a rotating BTZ spacetime.
The BTZ spacetime has two intrinsic radii, given
by the zeros 
of the metric component $g^{rr}$
in Eq.~(\ref{4::exterior}). They are 
the gravitational radius $r_+$
and the Cauchy radius $r_-$, 
which have the expressions
\begin{equation}
r_{+} = 2l \sqrt{G m+ \sqrt{G^2 m^2- \dfrac{G^2J^2
}{l^2}}} \,,  \label{gravr}
\end{equation}
\begin{equation}
r_{-} = 2l \sqrt{G m - \sqrt{G^2 m^2- \dfrac{G^2J^2
}{l^2}}} \,.  \label{cauchyr}
\end{equation}
The inequality $r_-\leq r_+$ always 
holds, with the equality  $r_-=r_+$
being equivalent to $J=ml$.
Inverting Eqs.~(\ref{gravr}) and (\ref{cauchyr})
gives $m$ and $J$ in terms of $r_+$ and 
$r_-$,
\begin{equation}
m= 
\frac{r_+^2+r_-^2}{8Gl^2}
\,,  \label{massadm}
\end{equation}
\begin{equation}
J= \frac{r_+r_-}{4Gl}
\,.  \label{jadm}
\end{equation}

Up to now we have put no constraints on the 
angular momentum $J$ of the outside spacetime. 
There are three cases we should discuss. 
First, $J<ml$, which is equivalent to $r_+>r_-$,
in which case the shell is underspinning.
Second, 
$J=ml$, which is equivalent to $r_+=r_-$,
and the outside spacetime is extremal 
BTZ and the shell is called extremal.
Third, $J>ml$, in this case 
both $r_+$ and $r_-$ are imaginary, 
and the spacetime and the shell are overspinning.
For the first two cases 
$r_+\geq r_-$, we
impose the mechanical constraint 
that the shell must be outside
its own gravitational radius, i.e., 
there are no trapped surfaces or horizons in the spacetime.
This means that the condition
\begin{equation}
R\geq r_{+}
\label{constraint1}
\end{equation}
must hold. For the other case, 
$J>ml$, the spacetime is overspinning and the shell 
can be taken to $R=0$ in which case is a naked singularity.
Thus, for this case one has to impose 
simply $R\geq0$. We will not discuss
further this latter case as it is too simple and the analysis
done so far is enough. So in the  rest of the paper 
we assume 
\begin{equation}
 r_{+}\geq r_{-}\,.
\label{constraint2}
\end{equation}

\subsection{The metric and matter quantities 
in terms of $r_+$ and $r_-$}

We have found all the relevant expressions 
for the ring-shell spacetime and have displayed 
the two intrinsic important radii of the spacetime,
$r_+$ and $r_-$. It is 
thus convenient to write the metric and the 
matter quantities  in terms of $r_+$ and $r_-$,
instead of $m$ and $J$. 
We thus use 
Eqs.~(\ref{gravr})-(\ref{jadm}).
From 
Eq.~(\ref{4::exterior}),
the exterior BTZ metric can be written as 
\begin{align}\label{4::exteriorr+}
ds_o^2 = &-\dfrac{1}{l^2r^2} 
\left[
\left(r^2-r_+^2\right)
\left(r^2-r_-^2\right)
\right]\,dt_o^2 
+ \frac{l^2r^2\,dr^2}{
\left(r^2-r_+^2\right)
\left(r^2-r_-^2\right)
} 
\nonumber\\
&
+ r^2 \left(d\phi-\frac{r_+r_-}{lr^2}dt_o\right)^2
\,,\quad r> R\,.&
\end{align}
The angular velocity $\Omega$ that makes the exterior 
metric diagonal at $\Sigma$, i.e., 
Eq.~(\ref{newpsi}),  can be written as
\begin{equation}
\Omega = \dfrac{r_+r_-}{lR^2} \label{Omegar+} \, .
\end{equation}
In coordinates $(t_o,\psi)$,
the induced metric at $\Sigma$ is then
\begin{equation} \label{LEOia2r+}
ds_{\Sigma}^2 =  -\dfrac{1}{l^2
R^2} 
\left[
\left(
R^2-r_+^2\right)
\left(
R^2-r_-^2\right)
\right]\,dt_o^2  + R^2 d\psi^2 \,,\quad r= R\,.
\end{equation}
In the rest frame of the shell, the components of the energy-momentum
tensor $S^a_b$, Eqs.~(\ref{lambdafluid})-(\ref{flu})
are given by
\begin{eqnarray}
\lambda&=&\frac{1}{8\pi Gl}
\Big[
1-\frac{1}{R^2}\sqrt{(R^2-r_+^2)(R^2-r_-^2)}\Big],
\label{lambdar+}\\
p&=&\frac{1}{8\pi G l}
\Big[
\frac{R^4-r_+^2r_-^2}{R^2\sqrt{(R^2-r_+^2)(R^2-r_-^2)}}
-1
\Big]\,,
\label{pr+}\\
j&=&\frac{r_+ r_-}{8\pi G l R}\,.
\label{jr+}
\end{eqnarray}

\section{The fluid seen as a perfect fluid}
\label{perfectfluid}

\subsection{The perfect fluid energy density
and pressure, and the angular velocity
of the reference frame that detects
a perfect fluid}

We now want to pass to a reference 
frame where the energy-momentum
tensor of the matter has the form of 
a perfect fluid energy-momentum
tensor. We will see that this is possible. 
Thus, we want that
the energy-momentum tensor takes the form 
\begin{equation}
{S}^{ab}={\bar \lambda}\, u^a u^b + {\bar p}
\left( u^a u^b + h^{ab}\right)\,,
\label{empfl}
\end{equation}
where $h_{ab}$ is the metric 
on the shell given in Eq.~(\ref{4::R}), 
$u^a$ is some velocity field on the shell,
and perfect fluid quantities are written
as barred quantities, $\bar \lambda$ 
is the perfect fluid energy density
and $\bar p$ its pressure.

Due to the circular symmetry and the fact that there are
only two components for the velocity $u^a$, one can write 
it as $u^a = \gamma (
\tau^a + {\bar \omega} \psi^a)$, for some 
angular velocity $\bar \omega$, and 
$\tau^a=\frac{\partial x_\Sigma^a}{\partial \tau}$,
$\psi^a=\frac{\partial x_\Sigma^a}{\partial \psi}$. 
Then the normalization
condition gives $\gamma=1/\sqrt{1-{\bar \omega}^2R^2}$
and so 
\begin{equation}
u^a = \frac{1}{\sqrt{1-{\bar \omega}^2R^2}} (
\tau^a + {\bar \omega} \psi^a) \,.
\end{equation}
Since $u^au_a=-1$ we have from Eq.~(\ref{empfl}) that
\begin{equation}
S^{a}_{b}\,u^b=-{\bar \lambda}\, u^a\,,
\label{eigen}
\end{equation}
meaning that the velocity field $u^a$
is an eigenvector of $S^{a}_{b}$
with eigenvalue $-{\bar \lambda}$. This 
yields two equations which enable to calculate 
${\bar \lambda}$ and ${\bar \omega}$. 
Indeed, Eq.~(\ref{eigen}) yields 
$\bar \lambda=-\frac{S^{\tau}_{\tau}+
R^2{\bar \omega}^2S^{\psi}_{\psi}}
{1+R^2{\bar \omega}^2}$ 
and $\frac{{\bar \omega}}{1+\bar\omega^2R^2}=
\frac{-S^{\psi}_{\tau}}{-S^{\tau}_{\tau}+S^{\psi}_{\psi}}$.
Having ${\bar \lambda}$ and 
$u^a$ one finds the pressure ${\bar p}$ by 
projecting 
${S}^{ab}$ into the direction 
orthogonal to $u^a$, i.e., 
\begin{equation}
{\bar p}=\left(h_{ab}+u_au_b\right){S}^{ab}\,.
\label{pperffl}
\end{equation}
Then using Eqs.~(\ref{eigen})-(\ref{pperffl})
together with  previous equations
one gets
\begin{eqnarray}
{\bar \lambda} &=&
\dfrac{1}{8 \pi G l} 
\left(
1-\sqrt{\frac{R^2-r_+^2}{R^2-r_-^2}}
\right)\,,
\label{lambdaperf}
\\
{\bar p}&=& \dfrac{1}{8 \pi Gl }
\left(
\sqrt{\frac{R^2-r_-^2}{R^2-r_+^2}}-1
\right)\,,
\label{pressurebtz}
\\
\bar \omega &=&\frac{r_-}{r_+ R}\sqrt{\frac{R^2-r_+^2}{R^2-r_-^2}}\,. 
\label{4omega}
\end{eqnarray}
Note that $\bar \omega$ is 
the angular velocity of
the reference frame which detects a perfect fluid
relative 
to the shell's proper frame, i.e., 
$\frac{d\psi}{d\tau}=\bar\omega$.
The derivation presented so far follows 
\cite{Poisson}$\,$.
For an alternative derivation see
Appendix.

\section{Effects due to rotation: Machian effects and 
dragging of the inertial frames}
\label{sevrotations}

\subsection{The several rotations and angular velocities}

We can now explore some effects due to rotation.
The BTZ metric given by  Eq.~(\ref{4::exterior}), 
appropriate for the vacuum 
region outside of the shell, tends 
to the AdS metric at infinity. 
Pure AdS metric is a nonrotating metric.
Thus, infinity 
is the standard of a nonrotating frame, the 
AdS frame, or the fixed stars frame
in Machian language. 
The empty metric given by  Eq.~(\ref{4::interior}), 
appropriate for the vacuum 
region inside of the shell, is in a coordinate
system that is rotating with respect to infinity. 
Indeed, from Eq.~(\ref{4::exterior}), 
one deduces that the interior region, 
including the interior neighborhood of the 
ring, has the property that lines with
constant $\psi$ move 
with respect to $t_o$, the global 
AdS time at infinity,
with angular velocity $d\phi/dt_o=\Omega$,
where $\Omega$ is given in 
Eq.~(\ref{Omegar+}) (see also
Eq.~(\ref{Omega})).
Now, since the inside is the empty 
space metric, lines with constant $\psi$ 
represent inertial frame lines 
for the interior spacetime that 
rotate with $\Omega$
with respect
to AdS infinity.

Moreover, the angular velocity $\bar \omega$,
found previously as the angular velocity 
of the frame that detects perfect
fluid quantities, 
is measured in the frame 
that rotates with $\Omega$ relative to 
infinity, the frame that 
has proper time $\tau$. 
Then, the corresponding angular velocity $\omega$ 
measured in the BTZ global time $t_o$ 
is 
$\omega
=\frac{d\psi}{dt_o}
=\frac{d\tau}{dt_o}\frac{d\psi}{d\tau}
=\frac{\sqrt{(R^2-r_+^2)(R^2-r_-^2)}}{l R}\bar \omega
=\frac{r_-}{l r_+}- \frac{r_+r_-}{l R^2}$, i.e., 
\begin{eqnarray}
\omega
=\frac{1}{l}\frac{r_-}{r_+}-\frac{1}{l}\frac{r_-}{r_+}
\frac{r_+^2}{R^2}\,.
\label{omegapure}
\end{eqnarray}
This  
$\omega$ is 
the angular velocity of
the reference frame which detects a perfect fluid
relative 
to the shell's proper frame redshifted 
to the global time $t_o$. 
It is always positive
since $R\geq r_+$.

In addition, we can calculate 
the angular velocity
of the reference frame at the shell that detects
a perfect fluid relative to infinity $t_o$.
Call this angular velocity  $\omega_\infty$.
Then, the angular velocity $\omega_\infty$
of the reference frame at the shell that detects
a perfect fluid relative to infinity $t_o$, is 
clearly the sum of the angular velocity $\omega$
of the reference frame that detects
a perfect fluid relative to the proper 
rest frame with time $\tau$
redshifted to the
global 
time $t_o$, 
plus the angular velocity 
of the shell relative to 
the global 
time  $t_o$, i.e., 
\begin{equation}
\omega_\infty=\frac{d\phi}{dt}=
\frac{d\psi}{dt} + \Omega = 
\omega + \Omega
\,.
\label{sumofom}
\end{equation}
The angular velocity $\omega_\infty$
of the reference frame at the shell that detects
a perfect fluid relative to 
a nonrotaing frame at infinity with time $t_o$,
is then 
\begin{equation}
\omega_\infty = \dfrac{1}{l}\dfrac{r_-}{r_+} \,,
\label{omegainf}
\end{equation}
where we have substituted 
Eqs.~(\ref{Omegar+}) and (\ref{omegapure})
in Eq.~(\ref{sumofom}).
Eq.~(\ref{omegainf}) does not depend on $R$, thus
$\omega_\infty$ is independent of the shell, 
it is an intrinsic property of the
spacetime.
This does not occur in 
the corresponding 3+1 dimensional spacetimes
\cite{Poisson}$\,$.

Remarkably, the angular velocity 
$\omega_\infty$ given in 
Eq.~(\ref{omegainf})
coincides with the horizon's angular velocity
of the BTZ black hole. 
Indeed, the pure BTZ spacetime, given 
by Eq.~(\ref{4::exterior}) for $0\leq r<\infty$,
possesses a null Killing vector normal to the 
horizon given by $n^\alpha=t_o^\alpha+
\omega_+\phi^\alpha$, where $\omega_+$ is 
the horizon angular velocity
\cite{btz1,btz2}$\,$. 
The condition $n^\alpha n_\alpha=0$
gives $g_{00}+2\omega_+g_{0\phi}
+\omega_+^2g_{\phi\phi}=0$, i.e., using Eq.~(\ref{4::exterior}),
$\frac{16G^2J^2}{r_+^2}-8\omega_+{GJ}
+\omega_+^2r_+^2=0$. The solution is
$\omega_+=\frac{4GJ}{r_+^2}$, i.e., 
\begin{equation}
\omega_+ = \dfrac{1}{l}\dfrac{r_-}{r_+}  \,.
\end{equation}
Thus, the angular velocity of the special 
reference frame at the shell 
$\omega_\infty$ has the same expression 
as the angular velocity of the 
BTZ pure vacuum horizon $\omega_+$ as seen 
by observers at 
infinity.

\subsection{Machian effects and the
dragging of the inertial frames}

We can continue to explore the effects due to 
the rotation of the ring.
As we have discussed, any inner line
or interior observer with 
$\psi$ constant is moving 
with respect to $t_o$ at infinity 
with angular velocity $d\phi/dt_o=\Omega$.
Now, observers with constant $\psi$ 
are inertial observers for the 
empty interior metric
and these observers rotate with respect
to AdS infinity with $\Omega$. 
Let us call the angular velocity
of the interior observers as $\Omega_{\rm in}$
with  $\Omega_{\rm in}=\Omega$.
This $\Omega_{\rm in}$
is caused 
by the presence of the rotation of the shell, 
and it is called
the dragging of inertial frames. 
From Eq.~(\ref{Omegar+})
we have, 
\begin{equation}
\Omega_{\rm in}=\frac{1}{l}\frac{r_-}{r_+}
\frac{r_+^2}{R^2} \,.
\end{equation}
We can compare this angular velocity 
of interior observers with the 
angular velocity of the shell
$\omega_\infty$ given in 
Eq.~(\ref{omegainf}). 
The ratio between the two angular velocities
$\Omega_{\rm in}/\omega_\infty$ is then
\begin{equation}
\frac{\Omega_{\rm in}}{\omega_\infty}=
\frac{r_+^2}{R^2}\,.
\end{equation}
For $R$ large the inside observers rotate
with a small fraction of the shell's 
reference frame in question.
For $R=r_+$, i.e., when the shell 
approaches its own gravitational radius, 
the inside observers corotate
with the shell 
$\frac{\Omega_{\rm in}}{\omega_\infty}=1$, 
indicating a very
strong dragging effect and a prominent 
example of Mach's principle.

The rotating effect on the inside 
region relative to infinity caused
by the presence of a rotating shell
as an example of 
the dragging of inertial frames
effect is well known in 3+1
general relativity, see e.g. \cite{Poisson}$\,$. 
We see that in 2+1 dimensions
this phenomenon also occurs.

\section{Energy Conditions}
\label{energyconditions}

\subsection{The weak energy condition}
In Eq.~(\ref{constraint1})
we have imposed 
that the shell's radius is always 
larger than the gravitational radius, $R\geq r_{+}$. 
We discuss here the weak and the dominant 
energy conditions.
The weak energy condition is automatically satisfied since we have
$\lambda$ and $p$ non-negative.

\subsection{The dominant energy condition}

The dominant energy condition can be discussed in 
the the  frame that detects the matter 
as a perfect fluid, where the 
quantities
$\bar \lambda$ and $\bar p$
are the relevant ones. In this frame the
dominant energy condition states that
\begin{equation}
\bar p\leq \bar \lambda\,.
\label{dom}
\end{equation}
From Eqs.~(\ref{lambdaperf}) and (\ref{pressurebtz}) this means
\begin{equation}
R^2-r_-^2\leq R^2-r_+^2\,.
\label{dom2}
\end{equation}

So, first, 
Eq.~(\ref{dom2}) can be obeyed 
when $R\to\infty$. In this case 
one has from 
Eqs.~(\ref{lambdaperf}) and (\ref{pressurebtz})
that 
${\bar \lambda}=\frac{1}{16\pi Gl}
\frac{r_+^2-r_-^2}{R^2}$
and 
${\bar p}=\frac{1}{16\pi Gl}
\frac{r_+^2-r_-^2}{R^2}$. 
Now, from Eq.~(\ref{menergy})
one has that for large $R$ that 
$m=\frac{R}{l}M$. So in this limit
$\bar \lambda=\frac{M}{2\pi R}$,  
$\bar p=\frac{M}{2\pi R}$, and 
$\bar p=\bar\lambda$. 
For $M$ constant the matter disappears  in the 
$R\to \infty$ limit. However, it can be the case that 
$M$ grows proportional to $R$, so that $\bar \lambda$ is constant.
In this case the limit $R\to \infty$ is also well defined 
and the shell obeys the dominant energy condition
since $\bar p=\bar \lambda$.

Second, Eq.~(\ref{dom2}) is satisfied 
when $r_+=r_-$ (or $J=ml$), i.e., the
outer BTZ spacetime is extremal and 
the observers that 
see a perfect fluid  are rotating  at the extremality
limit, i.e.,
at the speed of light, $R\bar \omega=1$,
see Eq.~(\ref{4omega}).
In this case, for these observers 
the matter in the shell becomes massless as
expected and so ${\bar \lambda}=0$ as one can check from
Eq.~(\ref{lambdaperf}). 
Now, the pressure for these observer also vanish, see
Eq.~(\ref{pressurebtz}). This can be interpreted 
considering that the AdS attraction due to the negative 
cosmological constant is now purely balanced by some kind 
of centrifugal force due to this maximum rotating speed,
making the pressure going to zero. 
In the limit one has $\bar p=\bar\lambda$ 
and so the dominant energy conditon is satisfied for any $R$. 
For all other settings the dominant energy condition
is violated. This is due to the
fact that the spacetime is AdS, with a negative 
cosmological constant, which as seen as a perfect fluid
does not obey the dominan energy condition.

\section{Conclusions}
\label{conclusions}

We have studied the dynamics of a 
rotating thin matter ring shell in 
(2+1)-dimensional spacetime
with a negative cosmological constant. 
The outside metric is the BTZ metric, 
asymptotically AdS, and the inside
metric is the empty spacetime metric
with a negative cosmological constant.
We
obtained
the shell's rest energy density, 
the pressure and angular momentu
flux density by using the
junction conditions. Due to
rotation, the shell's energy-momentum tensor was not in the form
of a perfect fluid. 
We then
attempted to write the energy-momentum tensor in a perfect fluid form 
and obtained the thin shell's  energy density, pressure and
angular velocity for the frame where one sees the thin shell as a
perfect fluid with no angular momentum flux.
We have found that Machian effects occur in 2+1 dimensions
and that frame dragging is present in these 
spacetimes.
We analyzed the weak and dominant energy conditions of the
system. The dominant energy condition
implies that there are only two valid configurations: the 
$R\to \infty$ case with its subcases, the trivial
$M=0$ shell and the nontrivial
$M=2\pi \lambda R$ shell, 
and the extremal case $r_{+} \rightarrow r_{-}$
(or $J=ml$).

\section*{Appendix: The energy-momentum tensor in a perfect fluid
form, another derivation}
\label{ap}

Here we derive properties of the matter in a frame that
sees it as a perfect fluid.
The intrinsic metric on the shell is given by
\begin{eqnarray}
ds^2=-d\tau^2+R^2d\psi^2.
\label{ms}
\end{eqnarray}
In the rest frame of the shell, 
the energy-momentum tensor $S^a_b$
is given by 
\begin{eqnarray}
\lambda&=&\frac{1}{8\pi G l}
 \Big(
1-\frac{1}{R^2}\sqrt{(R^2-r_+^2)(R^2-r_-^2)}\Big),
\\
p&=&\frac{1}{8\pi G l}
\Big(
\frac{R^4-r_+^2r_-^2}{R^2\sqrt{(R^2-r_+^2)(R^2-r_-^2)}}
-1
\Big)\,,\\
j&=&\frac{r_+ r_-}{8\pi Gl R}\,.
\end{eqnarray}
Now the 
metric on the shell, Eq.~(\ref{ms}), is invariant under the boost
\begin{eqnarray}
\label{lorentz}
\bar \tau=
\gamma
\Big(
\tau-\bar \omega R^2 \psi
\Big),
\quad 
\bar \psi
=
\gamma
\Big(
\psi- \bar \omega \tau
\Big),
\label{boo}
\end{eqnarray}
with 
\begin{eqnarray}
\gamma\equiv\frac{1}{\sqrt{1-\bar \omega^2R^2}}\,.
\label{g1}
\end{eqnarray}
Indeed, using Eq.~(\ref{boo}) in (\ref{ms}) one gets,
\begin{eqnarray}
ds^2
=-d\bar\tau^2+R^2d\bar\psi^2. 
\end{eqnarray}
Defining $S^{\bar \tau}{}_{\bar\tau}= -\bar \lambda$, 
$S^{\bar \psi}{}_{\bar \psi}=\bar p$,
and 
$S^{\bar \tau}{}_{\bar\psi}=\bar j$,
and transforming $S^a_b$ appropriately,
one obtains
\begin{eqnarray}
\bar \lambda
&=&
\gamma^2
\Big(
\lambda
-2\bar \omega  j
+\bar \omega^2 R^2p
\Big),
\\
\bar p
&=&
\gamma^2
\Big(
  p
-2\bar \omega j
+\bar \omega^2R^2\lambda
\Big)\,,
\\
\bar j
&=&
\gamma^2
\Big\{
\Big(1+\bar \omega^2R^2\Big) j
-\bar \omega R^2\big( \lambda+ p\big)
\Big\}\,.
\end{eqnarray}
Imposing $\bar j=0$ fixes 
\begin{eqnarray}
\label{om}
\bar \lambda&=&\frac{1}{8\pi G l}
\Big(
1-\sqrt{\frac{R^2-r_+^2}{R^2-r_-^2}}
\Big),
\\
\bar p&=&\frac{1}{8\pi G l}
\Big(
\sqrt{\frac{R^2-r_-^2}{R^2-r_+^2}}
-1
\Big)\,,\\
\bar \omega&=&
\frac{r_-}{ R r_+}\sqrt{\frac{R^2-r_+^2}{R^2-r_-^2}}\,,
\end{eqnarray}
which agreed with the results in Sec. 3.1. 
Thus,
\begin{eqnarray}
\label{inverse_rel}
\lambda
&=&
\gamma^2
\Big(
\bar\lambda
+\bar \omega^2R^2 \bar p
\Big),
\\
p
&=&
\gamma^2
\Big(
 \bar p
+\omega^2R^2\bar\lambda
\Big)\label{inverse_relp}\,,\\
j
&=&
\bar \omega R^2
\gamma^2\big(\bar\lambda+ \bar p\big),
\label{inverse_relj}\,.
\end{eqnarray}
The set of Eqs.~\eqref{inverse_rel}-\eqref{inverse_relp}
can be written in a perfect fluid form
\begin{equation}
{S}^{a}_{b}={\bar \lambda}\, u^a u_b + {\bar p}
\left( u^a u_b + h^{a}_{b}\right)\,,
\label{fluid}
\end{equation}
with
\begin{eqnarray}
&&u^\tau=\gamma
, \quad 
u^\psi=\gamma\bar \omega,\quad
u_\tau=-\gamma,\quad
u_\psi=\gamma \bar\omega R^2,
\\
&&h_{\tau\tau}=-1,
\quad
h_{\psi\psi}=R^2, \quad h_{\tau\psi}=0\,.
\end{eqnarray}

\section*{Acknowledgments}

We thank Gon\c{c}alo Quinta and Jorge Rocha for conversations.  We
thank FCT-Portugal for financial support through Project
No.~PEst-OE/FIS/UI0099/2014.  FJL thanks FCT for financial support
through project Incentivo/FIS/UI0099/2014.  MM thanks FCT for grant
number SFRH/BPD/88299/2012.



\begin{thebibliography}{99}


\bibitem{btz1} M. Ba{\~{n}}ados, C. Teitelboim, and J. Zanelli, ``The
black hole in three dimensional spacetime'', Phys. Rev. Lett. {\bf
69}, 1849 (1992).

\bibitem{btz2} M. Ba{\~{n}}ados, M. Henneaux, C. Teitelboim, and
J. Zanelli, ``Geometry of the 2+1 black hole'', Phys. Rev. D {\bf 48},
1506 (1994); arXiv:gr-qc/9302012.


\bibitem{pelegsteif} Y. Peleg and A. R. Steif, ``Phase transition for
gravitationally collapsing dust shells in 2+1 dimensions'',
Phys. Rev. D {\bf 51}, 3992 (1995); arXiv:gr-qc/9412023.

\bibitem{mannohpark0} R. B. Mann and J. J. Oh, ``Gravitationally
collapsing shells in (2+1) dimensions'', Phys. Rev. D {\bf 74}, 124016
(2006); arXiv:gr-qc/0609094.

\bibitem{LemosQuintaBTZ} J. P. S. Lemos and G. M. Quinta, ``Entropy of
thin shells in a (2+1)-dimensional asymptotically AdS spacetime and
the BTZ black hole limit'', Phys. Rev. D {\bf 89}, 084051 (2014);
arXiv:1403.0579 [gr-qc].

\bibitem{mannohpark} R. B. Mann, J. J. Oh, and M.-I. Park, ``The role
of angular momentum and cosmic censorship in the (2+1)-dimensional
rotating shell collapse'', Phys. Rev. D {\bf 79}, 064005 (2008);
arXiv:0812.2297 [hep-th].

\bibitem{Poisson} E. Poisson, {\it A Relativist's Toolkit: The
Mathematics of Black-Hole Mechanics}, (Cambridge University Press,
Cambridge, 2004).




\end{thebibliography}
\end{document}